# Effect of piezoelectric polarization on phonon group velocity in Nitride Wurtzites

Bijay Kumar Sahoo

*Department of Physics, N.I.T, Raipur, India*
e-mail : bksahoo.phy@nitrr.ac.in
Tel. no: +91 95891 86713
Fax no: 0771 22 54 600

## Abstract

We have investigated the effect of piezoelectric (PZ) polarization property on group velocity of phonons in binary as well as in ternary wurtzite nitrides. It is found that with the presence of PZ polarization property, the phonon group velocity is modified. The change in phonon group velocity due to PZ polarization effect directly depends on piezoelectric tensor value. Using different piezoelectric tensor values recommended by different workers in the literature, percent change in group velocities of phonons has been estimated. The Debye temperatures and frequencies of binary nitrides GaN, AlN and InN are also calculated using the modified group velocities. For ternary nitrides $Al_xGa_{(1-x)}N$, $In_xGa_{(1-x)}N$ and $In_xAl_{(1-x)}N$, the phonon group velocities have been calculated as a functions of composition. A small positive bowing is observed in phonon group velocities of ternary alloys. Percent variations in phonon group velocities are also calculated for a straightforward comparison among ternary nitrides. The results are expected to show a change in phonon relaxation rates and thermal conductivity of III-nitrides when piezoelectric polarization property is taken into account.

## I. INTRODUCTION

The wurtzite nitride semiconductors (GaN, AlN, InN and their alloys) are promising materials for the next generation of high-power optoelectronics devices [1, 2]. One of the important issues in further development of wurtzite nitride is self-heating. Self-heating strongly affects the performance of the device [3, 4]. Recently, wurtzite nitride LEDs suffer from the problem of efficiency droop which resulted from the piezoelectric effect [5]. The material in the active layer and the substrate material generally determine the thermal resistance of the device structure. Thus, it is important to know the accurate value of the thermal conductivity, k of corresponding material.

Several groups have reported investigation of the thermal conductivity value in wurtzite nitride films [6, 7, 8, and 9]. The thermal conductivity k for GaN films grown by hydride vapor phase epitaxy (HVPE) is about 1.3 W/cm K at room temperature [6]. Florscu et al. [8] have determined thermal conductivity of GaN films fabricated by lateral epitaxial overgrowth is about 1.7 – 1.8 W/cm K using the scanning thermal microscopy technique. S. Adachi has reported the k value 1.95 W/cm K for GaN films [10]. Some samples with lower dislocation density the thermal conductivity value reaches k = 2.1 W/ cm.K. The theoretical studies of the thermal conductivity of bulk GaN, conducted by Zou and co-workers [11, 12, 13], on the basis of the Callaway - Klemens approach [14, 15], have reported k value from 3.36 to 5.40 W/cm K by taking two sets of material parameters for GaN. The relatively large difference between these two values of the thermal conductivity is entirely due to ambiguity in values of material parameters of GaN. In the same work they have also demonstrated the effect of impurities and dislocations on thermal conductivity. The piezoelectric polarization effect on thermal conductivity, so far in our knowledge has not been considered. As nitride wurtzites are strong piezoelectric polarization materials, the effect should be taken for study. In this work, we have calculated the group velocity of the phonons by adding piezoelectric polarization property. Using modified phonon group velocity, calculation is done for Debye temperature and frequency for both binary and ternary wurtzite nitrides.

The paper is organized in the following manner. In the next section a brief introduction about spontaneous and piezoelectric polarization phenomenon in wurtzite nitrides is discussed. The modified phonon group velocity due to polarization effect is presented in section III. In section IV, the modified phonon group velocities are calculated for binary as well as for ternary wirtzite nitrides. Conclusions are given in section V.

**II. SPONTANEOUS AND PIEZOELCTRIC POLARIZATION**

A unique feature of the wurtzite nitrides is strong PZ polarization effect. It has been theoretically calculated and experimentally confirmed. Both spontaneous polarization (SP) and piezoelectric polarization (PZ) present in the wurtzite nitrides influence the optical, electrical and thermal properties and thus, significant consequence on device design [15]. Polarization induces sheet charges and electric fields values as high as 1 MV/cm [16] in these materials. The effect can be used constructively in device design to enhance carrier collection, ohmic resistance reduction and bending bands to reduce surface recombination.

Polarization is defined by $D = \varepsilon_0 \varepsilon_r E + P$, where, D is electric flux density (C/m$^2$), E is electric field (V/m), P is net polarization (C/m$^2$), $\varepsilon_0$ is vacuum permittivity, $\varepsilon_r$ is relative permittivity. The net polarization is composed of two parts: (i) spontaneous, which is intrinsic to a material, and (ii) piezoelectric, which is strain-induced, $P = P_{SP} + P_{PZ}$ where $P_{SP}$ is spontaneous polarization (C/m$^2$) and $P_{PZ}$ is piezoelectric polarization (C/m$^2$). The spontaneous polarization along the z (or c) axis of the wurtzite crystal is $\vec{P}_{SP} = P_{SP}\, z$. The piezoelectric polarization can be calculated by $P_{PZ} = \in_x e_{31} + \in_y e_{32} + \in_z e_{33}$, where $\in_x, \in_y, \in_z$ are strain components in the x, y, z direction, and $e_{31}, e_{32}, e_{33}$ are piezoelectric constants. The strain along the z-axis is $\in_z = (c - c_0)/c_0$ and the in-plane strain $\in_x = \in_y = (a - a_0)/a_0$ is assumed to be isotropic. The relation between $\in_x$ and $\in_z$ is $\in_z = -2\,(C_{13}/C_{33})\in_x$, where $C_{13}, C_{33}$ are elastic constants. Table 1 [17] shows the values of elastic constants of wurtzite nitrides (GaN etc.). Using the relation $\in_x$ and $\in_z$, the piezoelectric polarization in the direction of the c – axis becomes $P_{PZ} = 2(a - a_0)/a_0\,[\,e_{31} - e_{33}(C_{13}/C_{33})\,]$. The spontaneous and piezoelectric polarization values of wurtzite nitrides (GaN etc.) are presented in Table 2 [17, 18, and 19].

### III. MODIFIED PHONON VELOCITY

The coupling between the elastic strain and electric field is characterized by the third rank piezoelectric tensor $e_{\alpha\beta\gamma}$, according to the relation [20]

$$\sigma_{\alpha\beta} = \sum_{\gamma\delta} C_{\alpha\beta\gamma\delta} \frac{\partial u_\gamma}{\partial x_\delta} - \sum_\gamma e_{\gamma\alpha\beta}\, E_\gamma \qquad (1)$$

Where $\sigma_{\alpha\beta}$ is the stress tensor, $C_{\alpha\beta\gamma\delta}$ is the fourth-rank elastic modulus Tensor, $u_\gamma$ is the particle displacement, and $E_\gamma$ is the electric field. The equations of motion take the form

$$\rho \frac{\partial^2 u_\alpha}{\partial t^2} = \sum_\beta \frac{\partial \sigma_{\alpha\beta}}{\partial x_\beta}$$

$$= \sum_{\beta\gamma\delta} C_{\alpha\beta\gamma\delta} \frac{\partial^2 u_\gamma}{\partial x_\beta \partial x_\delta} - \sum_{\beta\gamma} e_{\gamma\alpha\beta} \frac{\partial E_\gamma}{\partial x_\beta} \tag{2}$$

The constitutive relation involving the electric displacement $D_\alpha$ and the dielectric tensor $\varepsilon_{\alpha\beta}$ when piezoelectricity is present, is

$$D_\alpha = \sum_{\alpha\beta} e_{\alpha\beta\gamma} \frac{\partial u_\beta}{\partial x_\gamma} + \varepsilon_0 \sum_\beta \varepsilon_{\alpha\beta} E_\beta. \tag{3}$$

Combining this relation with the Maxwell equation

$$\nabla \cdot D = 0 \tag{4}$$

yields, $\sum_{\alpha\beta\gamma} e_{\alpha\beta\gamma} \dfrac{\partial^2 u_\beta}{\partial x_\alpha \partial x_\gamma} + \sum_{\alpha\beta} \varepsilon_0 \varepsilon_{\alpha\beta} E_\beta = 0$. (5)

which is to be solved simultaneously with Eq. (3). As a specific example, let us consider the wurtzite form of nitrides (GaN, AlN, InN) with its six fold axis parallel to the z-direction. For elastic waves polarized in the z- direction and propagating in the x- direction, there is a single element of the piezoelectric tensor to be considered, conventionally designated by $e_{15}$ and a single element of the dielectric tensor $\varepsilon_{xx}$. The equations of motion and the Maxwell equation reduces to

$$\rho \frac{\partial^2 u_z}{\partial t^2} = C_{44} \frac{\partial^2 u_z}{\partial x^2} - e_{15} \frac{\partial E_x}{\partial x} \tag{6}$$

$$0 = e_{15} \frac{\partial^2 u_z}{\partial x^2} + \varepsilon_0 \varepsilon_{xx} \frac{\partial E_x}{\partial x}, \tag{7}$$

where $C_{44}$ is the appropriate elastic modulus in the Voigt notation. Eliminating the electric field from the equation of motion, we obtain the wave equation

$$\rho \frac{\partial^2 u_z}{\partial t^2} = \overline{C}_{44} \frac{\partial^2 u_z}{\partial x^2}, \tag{8}$$

With $\overline{C}_{44}$ the effective elastic modulus specified by

$$\overline{C}_{44} = C_{44} + [e_{15}^2 / (\varepsilon_0 \varepsilon_{11})]. \tag{9}$$

The Dispersion relation is $\omega [\overline{C}_{44}/\rho]^{1/2} q = \upsilon q$ here $\upsilon$ is the speed of the wave. It is clear that the speed is enhanced by the piezoelectric polarization effect [20]. Without piezoelectric polarization effect the transverse velocity of phonon is $\upsilon_T = [C_{44}/\rho]^{1/2}$. When piezoelectric effect is considered the transverse velocity becomes $\upsilon_{T,P} = \sqrt{\left(C_{44} + (e_{15}^2/\varepsilon_0 \varepsilon_{11})\right)/\rho}$. In real crystals, the sound velocity $\upsilon(q)$ depends on the direction and magnitude of the phonon wave vector $q$, and is specific to a given phonon polarization type. According to Callaway's formulation [13, 14], a single-branch polarization-averaged velocity $\upsilon$ along a specified crystallographic direction is

$$\upsilon = \left[\frac{1}{3}\left(\frac{1}{\upsilon_{T,1}} + \frac{1}{\upsilon_{T,2}} + \frac{1}{\upsilon_L}\right)\right]^{-1} \qquad (11)$$

Where $\upsilon_L$ and $\upsilon_{T,1,2}$ are the longitudinal and transverse sound velocities, respectively. Along [0001] direction, two transverse branches are degenerate and have the same velocity given by $\upsilon_{T,1} = \upsilon_{T,2} = \upsilon_T = (\overline{C}_{44}/\rho)^{1/2}$; the extra term is due to piezoelectric polarization effect, the longitudinal velocity is defined as $\upsilon_L = (C_{33}/\rho)^{1/2}$. Here $C_{ij}$ are the elastic constants of the crystal. It should be noted that longitudinal velocity does not change due to polarization effect while transverse velocity is affected. The relation that defines the Debye frequency is $\nu_D = \upsilon (3N/4\pi V)^{1/3}$. Here, N is the number of atoms present in volume V of the crystal and $\upsilon$ is phonon group velocity. Debye temperature $\theta_D$ is defined as $\theta_D = h\nu_D/k_B$, where $\nu_D$ is the Debye frequency, $h$ is planck's constant and $k_B$ is Boltzmann's constant. The Debye frequency and temperature are changed because of change in phonon group velocities of III-nitrides.

## IV. RSULT AND DISCUSSION

The group velocity of phonons in III binary nitrides without considering PZ polarization property are first calculated using values of elastic constants from Table 1. With the presence of PZ polarization property, the effective elastic constant $\overline{C}_{44}$ of binary nitrides are then calculated using values of elastic constants and piezoelectric tensor elements from Table 2. For $e_{15}$ great differences are found: not even the sign is reported consistently in literatures [21]. Bernardini and Fiorentini [19] published values of the complete piezoelectric tensors for GaN, AlN, and InN. The values for InN reported by them are cited by Vurgaftman and Meyer in their review [18], because only a few values are available in the literature. Bernardini and Fiorentini derived their results by first-principles calculations and came up with positive values for $e_{15}$ for all three III-nitrides. According to reference [19] $e_{15}$ = +0.19 for GaN, +0.34 for AlN, +0.26 for InN. Using these values of $e_{15}$, we have calculated the effective elastic constants and then phonon group velocities of binary nitrides. These give percent variations in phonon group velocities equal to 0.18 % for GaN, 0.48 % for AlN and 0.42 % for InN. From reference [18], $e_{15}$= +0.33 for GaN, +0.42 for AlN, +0.26 for InN. All three values are positive. These give percent variation in phonon group velocities equal to 0.54 % for GaN, 0.74 % for AlN and 0.42 % for InN. Very recently, Romanov et al [17] published values for the piezoelectric tensor elements for GaN, InN, and AlN together with their strain calculations for arbitrarily grown pseudomorphic layers. According to reference [17], $e_{15}$ (C.m$^{-2}$) = -0.40 for GaN, -0.48 for AlN and -0.40 for InN. All the three values are negative. For InN, they used the same value as for GaN. These give percent variation in phonon group velocities equal to 0.80 % for GaN, 0.96 % for AlN and 0.99 % for InN. Table 3 shows the percent change in phonon group velocities for different recommended values of $e_{15}$. The corresponding change in Debye's frequencies and temperatures are shown in Tables 4 and 5 respectively. The Debye temperature for binary wurtzite nitrides GaN, AlN and InN without PZ polarization are 600, 1150 and 660 $^0$K respectively [22]. In addition, the percent changes in phonon group velocities for GaN, AlN and InN due to addition of PZ polarization property are shown in Fig1 for straightforward comparison. Guy and co-workers [23, 24] published experimental results for all piezoelectric tensor elements. They found $e_{15}$ > 0. The experimental determination of $e_{15}$ should at least yield a more reliable sign than the assumption used before. We therefore consider the reported positive value of $e_{15}$ as more accurate than the negative one of earlier works [22]. For our other calculations in this paper, we have used the values of $e_{15}$ recommended by Vurgaftman et al. That is $e_{15}$= +0.33 for GaN, +0.42 for AlN, +0.26 for InN, because these recommended values are intermediate between the most reliable theoretical and experimental values of $e_{15}$ [18]. To calculate phonon group velocities of ternary nitrides, we have explored the effective elastic constant $\overline{C}_{44}$, piezoelectric polarization tensor $e_{15}$, dielectric constant $\varepsilon_{11}$ and density $\rho$ of the ternary alloys as a function of composition $x$ by interpolating their corresponding binary values. The variation of effective elastic constant $\overline{C}_{44}$ and $e_{15}$ of ternary alloys with composition are shown in Fig 2 and Fig 3 respectively. From figs 2 and 3, it can be observed that $\overline{C}_{44}$ and $e_{15}$ vary linearly with composition i.e., they obey Vegard's law. However, the variations of phonon group velocities of the ternary alloys do not show a linear dependence on composition [19, 25]. This result is shown in Fig 4. It can be observed from fig 4 that, there is a small positive bowing in each curve; even if we have not used the bowing parameter in the calculations. To know maximum change in phonon group velocities with composition among the ternary alloys due to the addition of PZ polarization property, we have plotted the percent variation in phonon group velocities in Fig 5. Maximum change in phonon group velocity is observed in AlGaN and least variation is observed in case of InGaN. The Debye frequency and temperatures are directly proportional to phonon group velocities (section 3). Hence, the change in Debye frequencies and temperatures of ternary alloys as a function of composition will follow the similar type as that shown by phonon group velocities of ternary alloys.

## V. CONCLUSION

We have investigated theoretically the effect of PZ polarization property on phonon group velocity in binary as well as in ternary wurtzite nitrides. The effect of PZ polarization property on phonon group velocity directly depends on the values of PZ polarization tensor $e_{15}$. The percent variation in phonon group velocity, Debye frequency and Debye temperature of GaN, AlN and InN are calculated by considering three important values of PZ tensor recommended by Romanov et al, Vurgaftman et al and Bernardini et al. Using PZ tensor values recommended by Romanov et al, we found maximum percentage change in phonon group velocity, Debye frequency and Debye temperature in InN and least in GaN. However, for PZ tensor values recommended by Vurgaftman et al and Bernardini et al, maximum percent change is obtained in case of AlN and minimum is observed in case of GaN. The PZ tensor values recommended by Vurgaftman et al have been used for calculation of ternary alloys. The effective elastic constant $\overline{C}_{44}$ and piezoelectric polarization tensor $e_{15}$ of the ternary alloys vary linearly with composition. However, the variations of phonon group velocities of the ternary alloys do not show a linear dependence on composition. For straightforward comparison, percent variations in phonon group velocities of ternary alloys have been estimated. It is observed that the enhancement of phonon group velocity is prominent in AlGaN among the ternary nitrides and least enhancement in case of InGaN. The obtained results are expected to show a change in phonon relaxation rates and thermal conductivity of III-nitrides when piezoelectric polarization property is taken into account. This is reserved for future study.

## ACKNOWLEDGEMENTS

Author thanks Prof. S.M. Saini for fruitful discussion.

# Tables

**Table1**: Elastic constants of III-nitrides with wurtzite structure [ 17]

| Elastic constants | GaN | AlN | InN |
|---|---|---|---|
| $C_{11}$ (GPa) | 367 | 396 | 223 |
| $C_{12}$ (GPa) | 135 | 137 | 115 |
| $C_{13}$ (GPa) | 103 | 108 | 92 |
| $C_{33}$ (GPa) | 405 | 373 | 224 |
| $C_{44}$ (GPa) | 95 | 116 | 48 |

**Table 2:** Spontaneous polarization, piezoelectric polarization tensors and dielectric constants of AlN, GaN and InN [17, 18, and 19]

| Parameters | GaN | AlN | InN |
|---|---|---|---|
| $P_{SP}(C/m^2)$ | - 0.029 | - 0.081 | - 0.032 |
| $e_{33}$ $(C/m^2)$ | 0.73 | 1.55 | 0.73 |
| $e_{31}$ $(C/m^2)$ | -0.49 | - 0.58 | -0.49 |
| $e_{15}$ $(C/m^2)$ (Romanov *et al*) | - 0.40 | - 0.48 | - 0.40 |
| $e_{15}$ $(C/m^2)$ (Vurgaftman *et al*) | 0.33 | -0.42 | 0.26 |
| $e_{15}$ $(C/m^2)$ (Bernadini *et al*) | 0.19 | 0.34 | 0.26 |
| $\varepsilon_{11}$ $(=\varepsilon_{xx})$ | 9.5 | 9.0 | 15.3 |

**Table III:** Phonon group velocities (m/sec) in III - nitrides GaN, AlN and InN for different $e_{15}$ values.

| Literatures | Materials wurtzite | without PZ polarization | with PZ polarization | % Variation |
|---|---|---|---|---|
| $e_{15}$ value from Romanov et al [17] | GaN | 4746.11 | 4784.10 | 0.80 |
| | AlN | 6999.90 | 7067.60 | 0.96 |
| | InN | 3233.85 | 3265.88 | 0.99 |
| $e_{15}$ value from Vurgaftman et al [18] | GaN | 4746.12 | 4772.02 | 0.54 |
| | AlN | 6999.90 | 7051.86 | 0.74 |
| | InN | 3233.38 | 3247.45 | 0.42 |
| $e_{15}$ value from Bernardini et al [19] | GaN | 4746.11 | 4754.72 | 0.18 |
| | AlN | 6999.90 | 7034.06 | 0.48 |
| | InN | 3233.38 | 3247.45 | 0.42 |

**Table IV:** Debye frequencies (rad/sec) in III-nitrides (GaN, AlN and InN) for different $e_{15}$ values.

| Literatures | Materials Wurtzite | without PZ polarization | with PZ polarization | % Variation |
|---|---|---|---|---|
| $e_{15}$ value from Romanov et al [ ] | GaN | $1.249623 \times 10^{13}$ | $1.259625 \times 10^{13}$ | 0.80 |
| | AlN | $2.395110 \times 10^{13}$ | $2.418245 \times 10^{13}$ | 0.96 |
| | InN | $1.374585 \times 10^{13}$ | $1.388203 \times 10^{13}$ | 0.99 |
| $e_{15}$ value from Vurgaftman et al [ ] | GaN | $1.249623 \times 10^{13}$ | $1.256445 \times 10^{13}$ | 0.54 |

|  | | without PZ polarization | with PZ polarization | % Variation |
|---|---|---|---|---|
|  | AlN | $2.395110 \times 10^{13}$ | $2.412860 \times 10^{13}$ | 0.74 |
|  | InN | $1.374585 \times 10^{13}$ | $1.380367 \times 10^{13}$ | 0.42 |
| $e_{15}$ value from Bernardini *et al* [ ] | GaN | $1.249623 \times 10^{13}$ | $1.251892 \times 10^{13}$ | 0.18 |
|  | AlN | $2.395110 \times 10^{13}$ | $2.406769 \times 10^{13}$ | 0.48 |
|  | InN | $1.374585 \times 10^{13}$ | $1.380367 \times 10^{13}$ | 0.42 |

**Table V:** Debye temperatures ( $^0K$ ) in III-nitrides ( GaN, AlN and InN) for different $e_{15}$ values.

| Literatures | Materials Wurtzite | without PZ polarization | with PZ polarization | % Variation |
|---|---|---|---|---|
| $e_{15}$ value from Romanov *et al* [ ] | GaN | 600.00 | 604.80 | 0.80 |
|  | AlN | 1150.00 | 1161.10 | 0.96 |
|  | InN | 660.00 | 666.53 | 0.99 |
| $e_{15}$ value from Vurgaftman *et al* [ ] | GaN | 600.00 | 603.27 | 0.54 |
|  | AlN | 1150.00 | 1158.52 | 0.74 |
|  | InN | 660.00 | 662.77 | 0.42 |
| $e_{15}$ value from Bernardini *et al* [ ] | GaN | 600.00 | 601.08 | 0.18 |
|  | AlN | 1150.00 | 1155.59 | 0.48 |
|  | InN | 660.00 | 662.77 | 0.42 |

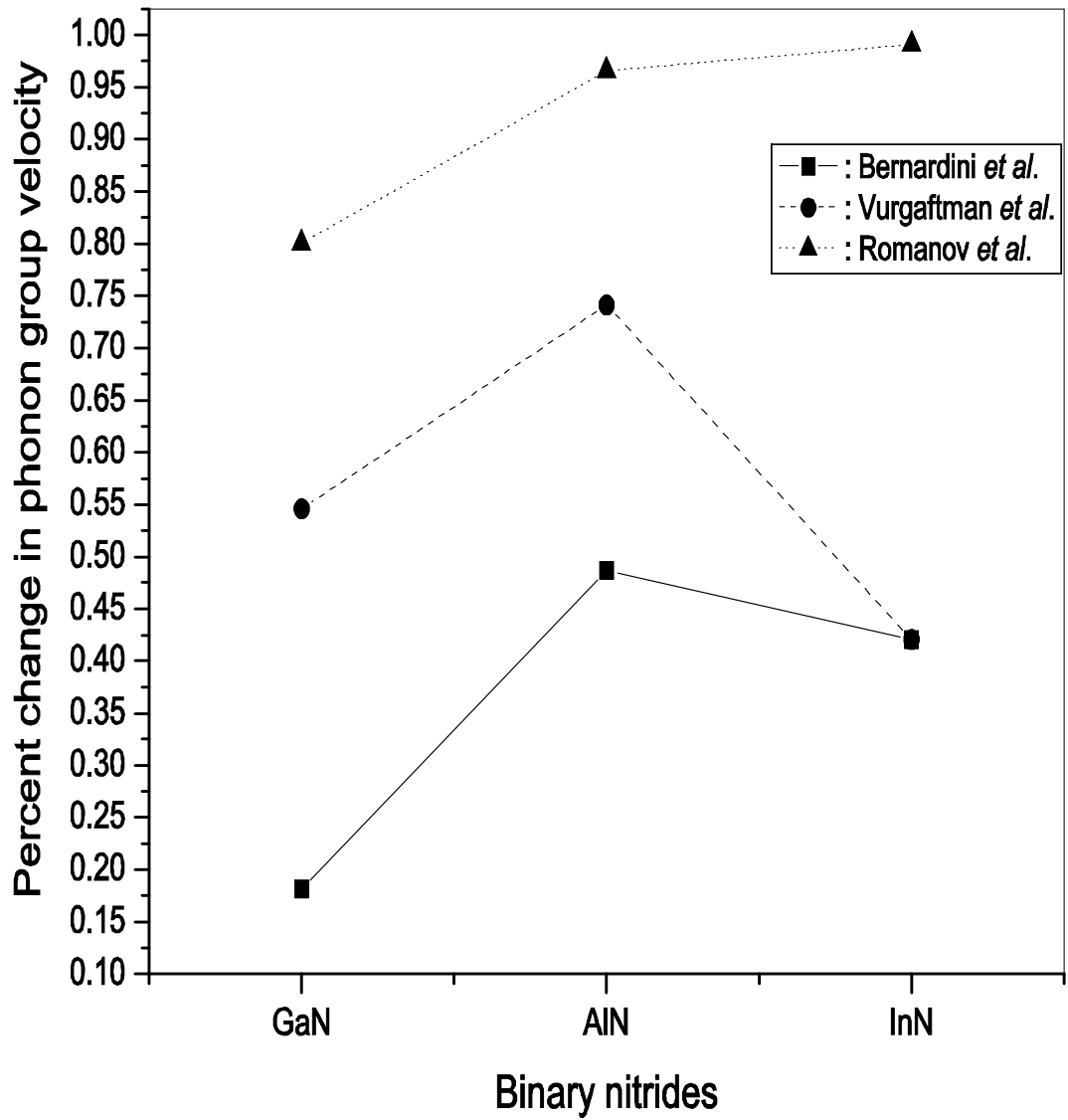

Fig.1 Percent change in phonon group velocities in binary III- nitrides for different $e_{15}$ values.

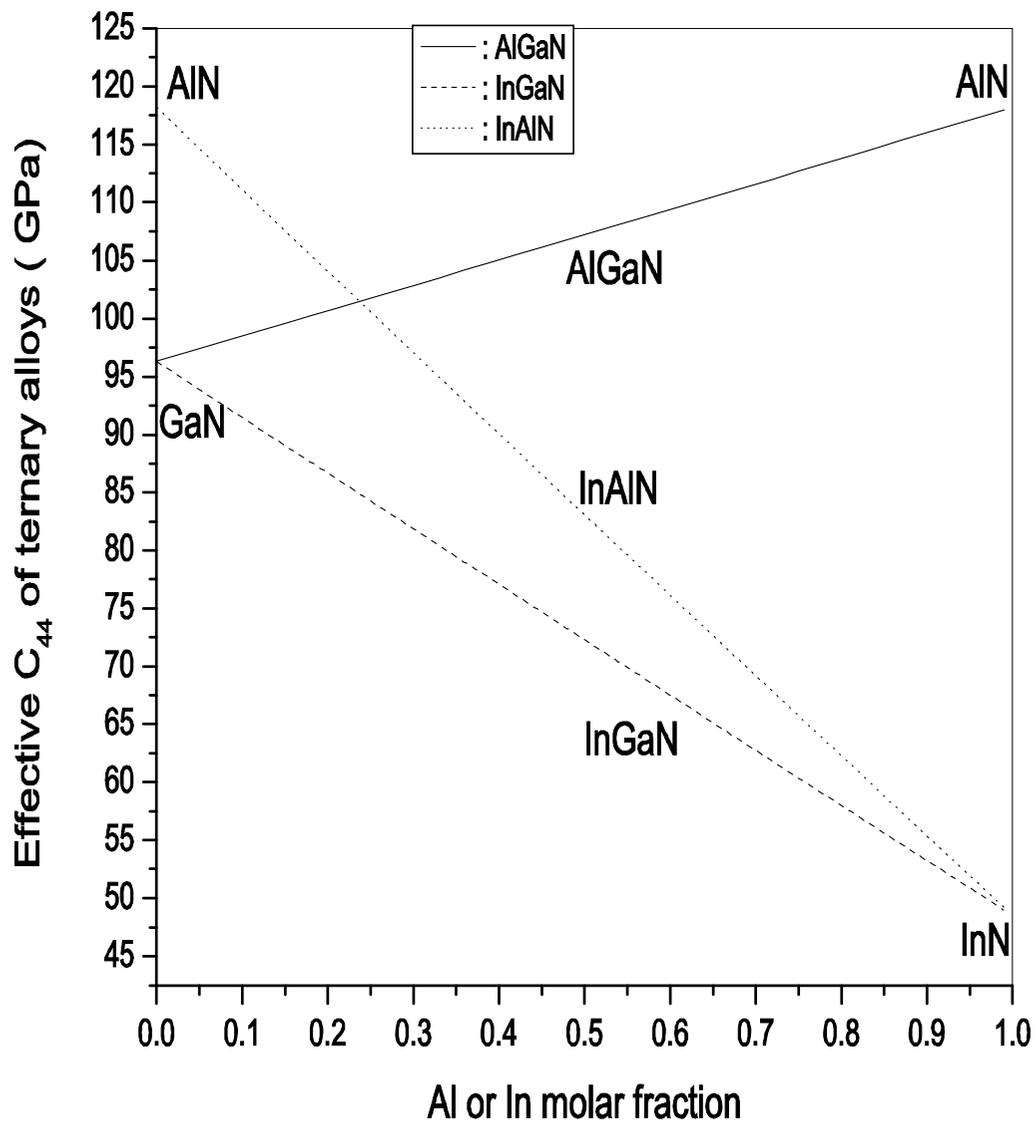

Fig.2 Elastic constants of ternary alloys as a function of composition

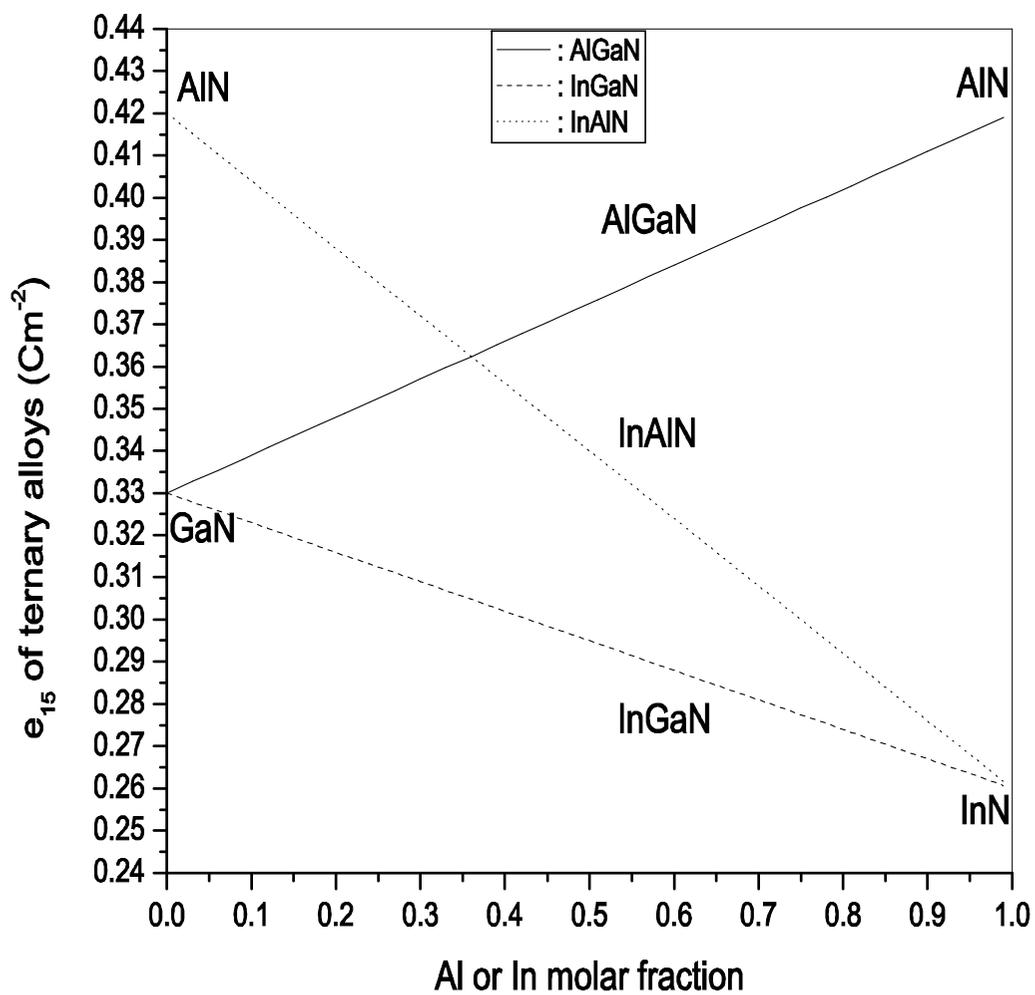

Fig.3 Piezoelectric tensor elements of ternary alloys as a function of composition

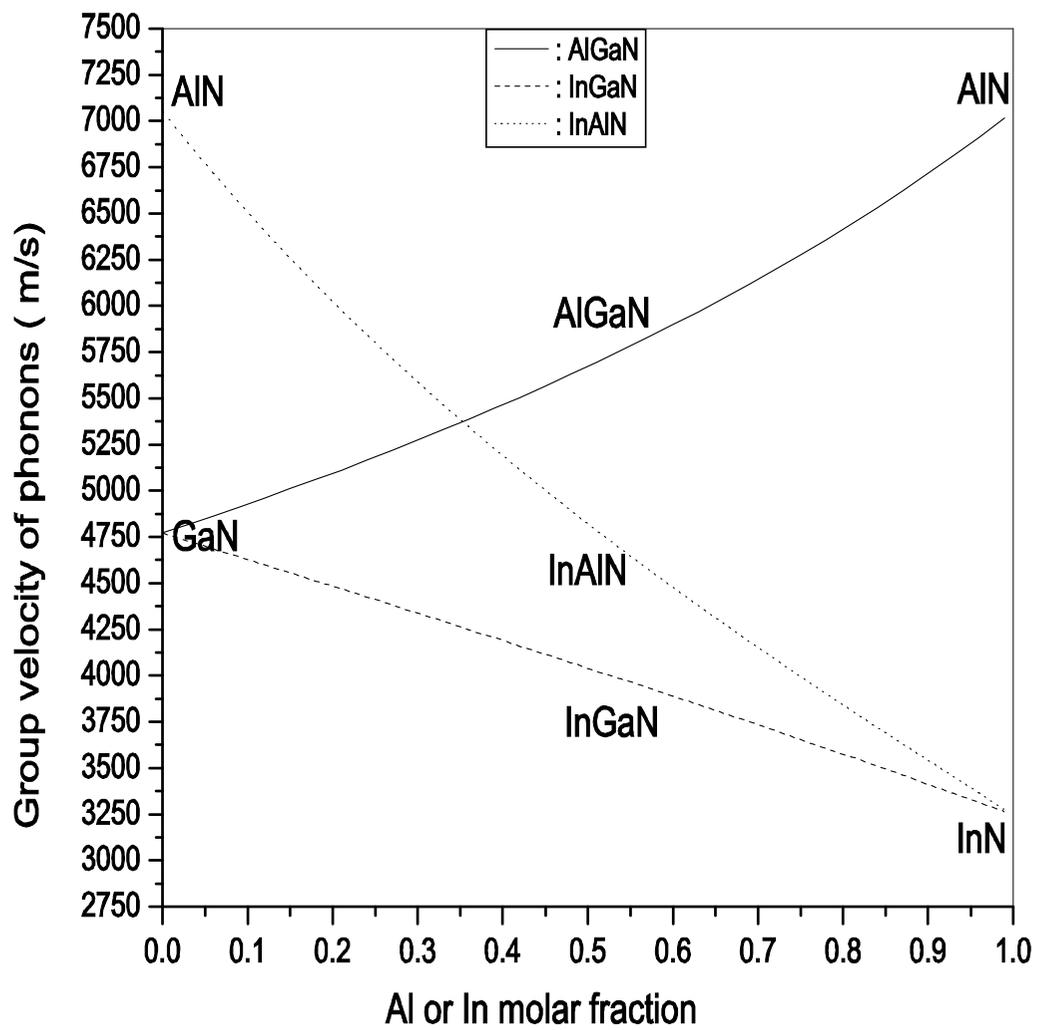

Fig. 4 Phonon group velocity of ternary alloys as a function of composition

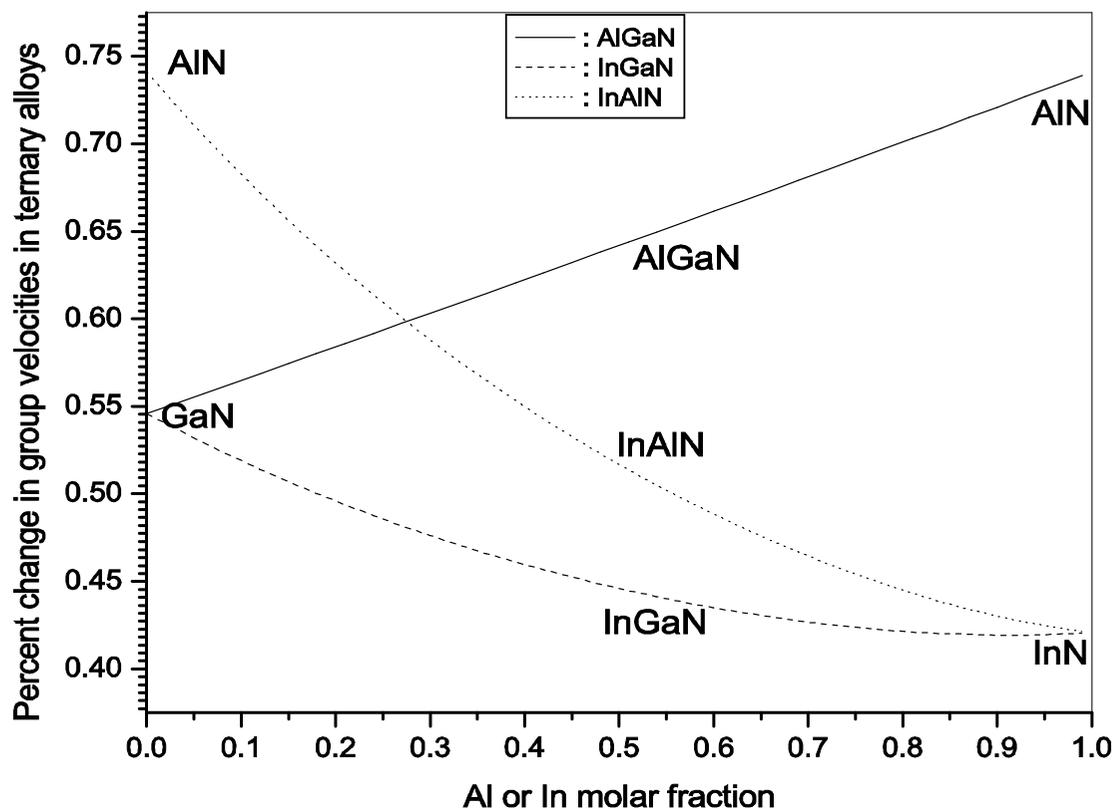

Fig. 5 Percent change of phonon group velocities in ternary alloys.